\documentclass[usenatbib,useAMS,usedcolumn]{mn2e}
\usepackage{graphicx}
\usepackage{subfigure}
\usepackage{url}
\usepackage{amssymb,amsmath}
\usepackage{multirow}

\usepackage{url}
\usepackage{nameref}
\usepackage[bookmarks=true,pdftitle={},pdfauthor={Roberto Raddi},pdfsubject={},pdfcreator={Roberto Raddi}, colorlinks=true, linkcolor=black, citecolor=black]{hyperref}
\usepackage[figure, figure*]{hypcap}
\usepackage{bbding}
\usepackage{breakurl}

\def\aap{A\&A}

\def\aj{AJ}
\def\apj{ApJ}
\def\apjs{ApJS}
\def\apjl{ApJ}

\def\mnras{MNRAS}
\def\memsai{MmSAI}
\def\ssr{SSRv}
\def\pasp{PASP}

\def\nat{Nature}
\def\zap{Zeitschrift f\"ur Astrophysik}

\def\Halpha{{\rm H}\alpha}

\title[Water-rich debris in a metal-polluted white dwarf]{
Likely detection of water-rich asteroid debris in a metal-polluted white dwarf}

\author[R. Raddi et al.]{R. Raddi$^1$\thanks{E-mail: r.raddi@warwick.ac.uk}, B.T. G\"ansicke$^1$, D. Koester$^2$, J. Farihi$^3$, J.J. Hermes$^1$,  
\newauthor S. Scaringi$^{4}$, E. Breedt$^1$, J. Girven$^1$\\
$^{1}$ Department of Physics, University of Warwick, Gibbet Hill Road, Coventry CV4 7AL, UK \\
$^{2}$ Institut f\"ur Theoretische Physik und Astrophysik, University of Kiel,
24098 Kiel, Germany\\
$^{3}$ University College London, Dept. of Physics \& Astronomy, London, WC1E 6BT, UK\\
$^{4}$ Max Planck Institut f\"ur Astrophysik, Karl-Schwarzschild-Str. 1, 85748, Garching, Germany}

\begin{document}

\date{Accepted 2015 March 26.  Received 2015 March 18; in original form 2015 February 18}

\pagerange{\pageref{firstpage}--\pageref{lastpage}} \pubyear{2015}

\maketitle

\label{firstpage}

\begin{abstract}
The cool white dwarf SDSS\,J124231.07+522626.6 exhibits photospheric absorption lines of 
8 distinct heavy elements in medium resolution optical spectra, notably including oxygen.
The $T_{\rm{eff}} = 13\,000$\,K atmosphere is helium-dominated,
but the convection zone contains significant amounts of hydrogen and oxygen.  
The four most common rock-forming elements (O, Mg, Si, and Fe) account for almost all the 
accreted mass, totalling at least $1.2 \times 10^{24}$\,g, similar to the mass of Ceres. 
The time-averaged accretion rate is $2\times10^{10}$\,g\,s$^{-1}$, one of the highest rates inferred among all known metal-polluted white dwarfs. 
We note a large oxygen excess, with respect to the most common metal oxides, suggesting that the white dwarf 
accreted planetary debris with a water content of $\approx 38$ per cent by mass. 
This star, together with GD\,61, GD\,16, and GD\,362, form a small group of outliers
from the known population of evolved planetary systems accreting predominantly dry, rocky debris.
This result strengthens the hypothesis that, integrated over the cooling ages of white dwarfs, accretion of water-rich 
debris from disrupted planetesimals may significantly contribute to the build-up of trace hydrogen observed in a large 
fraction of  helium-dominated white dwarf atmospheres.
\end{abstract}


\section{Introduction}
There is now evidence that at least 20--30 per cent of all white dwarfs with cooling ages $\gtrsim 100$\,Myr
have preserved parts of their planetary systems throughout the post-main sequence evolution 
\citep[][]{Zuckerman03,Zuckerman10,Koester14}, via detection of trace metals in their otherwise pure hydrogen
and helium atmospheres \citep[e.g.][]{Koester05a,Dufour07}.
These evolved planetary systems provide detailed information on the frequency, 
architecture, and chemical composition of extra solar planetary systems, once orbiting
early F-, A-, and late B-type stars of $\approx$\,1.2--3\,$M_\odot$.

Since the strong surface gravity of white dwarfs results in the radial stratification of their chemical constituents \citep[][]{Schatzman45,Schatzman47},
the heavy elements detected in the spectra of some white dwarfs 
must come from an external source \citep[][]{Dupuis92,Dupuis93a}.
The thermal emission from circumstellar dust has been detected in the infrared at 
about 35 white dwarfs \citep[][]{Reach05,Kilic06,Jura07,Farihi09,Bergfors14},
some of which show line emission from a gaseous component 
\citep[e.g.][]{Gaensicke06, Gaensicke08,Farihi12a,Melis12,Wilson14}.
Circumstellar debris discs at white dwarfs are thought to form after the tidal disruption of planetesimals \citep[][]{Debes02,Jura03,Veras14}.
\begin{figure*}
\centering
\includegraphics[width=\linewidth]{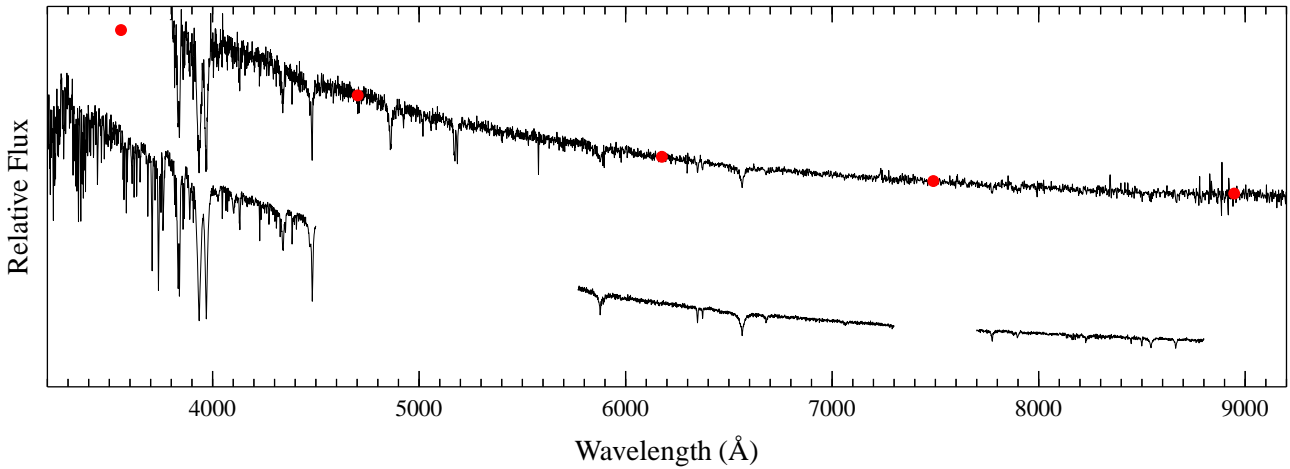}
\caption{Comparison between the SDSS (top) and WHT/ISIS (bottom) spectra of SDSS\,J1242, offset with respect to each other.
The wavelength coverage of the WHT/ISIS observations is more fragmentary, due to the adopted dual-arm configuration that
causes the gaps at 4500--5700\,\AA, and 7300--7700\,\AA. However, the data extend to shorter ultraviolet wavelengths and 
offer improved signal-to-noise ratio up to a factor of 5 times better than the SDSS spectrum. 
The SDSS broad-band photometry (in Table~\ref{t:photometry}) is also overplotted, as filled red dots, on to the SDSS spectrum.}
\label{f:sdss_isis_spectra}
\end{figure*}

To date only a handful of white dwarfs have been observed for detailed composition analysis, but for
these stars the bulk composition of the planetesimals causing metal-pollution has been inferred to be similar to that of
the inner Solar System, with the four most common rock-forming elements (O, Mg, Si, Fe) 
being the dominant species \citep[][see also \citealt{Jura14} for a review]{Zuckerman07, Klein10, Vennes10,Gaensicke12}.
Furthermore, the diverse composition exhibited by polluted white dwarfs, specifically with respect to the 
relative abundances of iron, siderophile, and refractory elements, also indicates melting, stripping and core differentiation
of extra solar asteroids \citep[e.g.][]{Melis11}. In most of the cases, the polluting debris appears to be notelet 
depleted of volatile elements such as H, C, N, and S, and generally drier than Solar System CI chondrites \citep[e.g.][]{Jura06,Klein11}. 

The presence of water in exoplanets is crucial in the context
of life outside the Solar System. White dwarfs offer a unique insight into the
ubiquity of water. \citet{Farihi13} presented the first unambiguous evidence for 
ongoing accretion of water-rich debris in GD\,61.
The helium-dominated atmosphere of this white dwarf contains a large mass of hydrogen, although not
unusual for stars of similar temperature, but the inferred mass of 
oxygen exceeds what is expected from the accretion of metal oxides.
This suggests that the parent body contained 26 per cent of water by mass.
Two other helium-dominated white dwarfs, GD\,16 and GD\,362, were also discussed in the context of accretion from water-rich debris
\citep[e.g.][]{Jura08,Jura09a, Jura10}, both stars showing circumstellar dust emission that reveals ongoing accretion.
However, while the atmospheres of these stars contain larger amounts of hydrogen ($\geq 10^{24}$\,g), along with traces of rock-forming metals,
the exact origin of hydrogen cannot be unambiguously linked to the circumstellar material.

Here, we present the discovery and follow-up observations of SDSS\,J124231.07+522626.6 (hereafter SDSS\,J1242),
a metal-polluted helium-dominated white dwarf with strong absorption lines of hydrogen, oxygen and seven other heavy elements.
We outline in Section~\ref{chap3.1} the model atmosphere analysis and the determination of element abundances.  In Section~\ref{chap3.2},
we analyse the infrared data, and comment on the non-detection of circumstellar dust emission. 
In Section~\ref{chap4}, we discuss the composition of the debris polluting the atmosphere of SDSS\,J1242 and 
highlight the large oxygen and hydrogen content, suggesting that SDSS\,J1242 accreted water-rich debris.
Finally, we summarise our findings in Section\,\ref{chap5}.

\section{Observations}
\label{chap2}

\subsection{Identification based on SDSS data}
Optical spectroscopy of SDSS\,J1242 was obtained by the Sloan
Digital Sky Survey \citep[SDSS;][]{York00} on 2002 April 15, and
released as part of Data Release 2 \citep[DR~2;][]{Abazajian04}. 
It was initially classified by \citet{Eisenstein06} as a subdwarf because of its narrow
Balmer and He\,{\sc ii} lines (Fig.~\ref{f:sdss_isis_spectra}). 
As part of our previous analysis of the Data Release 7 \citep[DR~7;][]{Abazajian09},
we retained this classification \citep[][]{Girven11}.
Upon closer inspection, we subsequently found the spectrum of SDSS\,J1242 to resemble those of other metal-polluted white dwarfs,
which appear as DAZ\footnote{We type SDSS\,J1242 as a DAZB white dwarf,
following the spectral classification scheme of \citet{Sion83}. D stands for degenerate star, while A, Z, and B 
indicate the presence of hydrogen, metallic, and helium lines ordered by line strength.} spectral types
under low resolution spectroscopy, and only high resolution spectroscopy showed
helium to be the dominant element in their atmospheres \citep[e.g. GD\,362, and GD\,16;][]{Zuckerman07, Koester05a}.
We therefore reclassify here SDSS\,J1242 as a white dwarf, and obtained
follow-up observations.

In Table~\ref{t:photometry}, we report the optical SDSS photometry and the ultraviolet
photometry from the Galaxy Evolution Explorer
\citep[{\em GALEX};][]{Morrissey07}.

\subsection{WHT/ISIS observations}
\label{chap2.1}
We observed SDSS\,J1242 with the Intermediate dispersion Spectrograph and Imaging System (ISIS) at the 4.2-m William Hershel Telescope (WHT) in La Palma. 

The WHT/ISIS spectra were taken on two occasions, in 2012 May 31, and 2012 December 14.
We used a 1 arcsec slit, the standard 5300 dichroic, and the R600B and R600R gratings in the blue and red arm, respectively.
The order-sorting filter GG495 was mounted on the red arm.
The blue arm covered 3100--4500\,\AA\ at a resolution of 1.8\,\AA.
We opted for two different configurations of the red arm, to cover the regions around
$\Halpha$ (5800--7300\,\AA) and Ca\,{\sc ii} triplet (7700--8800\,\AA) at a resolution of 2\,\AA.
With this setup, we obtained spectra bluer than the Balmer jump, which were not available from SDSS.

We observed in the blue arm for a total of 4.7\,hr, split roughly equally between the two runs. The $\Halpha$ region was exposed for
2.6\,hr, while the Ca\,{\sc ii} region was exposed for 2.3\,hr. 
Spectrophotometric standards and CuAr+CuNe arcs were acquired across the night, for the flux and wavelength
calibrations. We had clear sky and 1 arcsec seeing during both runs. 

We followed standard data-reduction procedures for removal of the bias, flat field correction, wavelength calibration, 
extraction of the spectrum, flux calibration and removal of telluric lines. 
All the reduction steps were undertaken with the {\sc pamela} \citep{Marsh89} and {\sc molly} packages\footnote{Both developed by T. R. Marsh. 
{\sc pamela} is part of the {\sc starlink} distribution at \url{http://starlink.jach.hawaii.edu/starlink}.\\
{\sc molly} is available at \url{http://www.warwick.ac.uk/go/trmarsh/software/}.}.

The coadded WHT/ISIS spectra and the SDSS spectrum are shown in Fig.~\ref{f:sdss_isis_spectra}. 
The quality of the WHT/ISIS spectra is superior with respect to the SDSS data in terms of wavelength coverage, resolution 
and signal-to-noise ratio (up to a factor of 5), which we measure to range between 20--40 in the red arm and 80--100 in the blue arm.
However, we find the WHT/ISIS spectrum to display 15 per cent more flux in the $u$-band with respect to the SDSS photometry,
probably arising from an imperfect flux calibration.

\subsection{Ground based near-infrared photometry}

Near-infrared $JHK_s$ photometry was obtained with the
Long-slit Intermediate Resolution Infrared Spectrograph \citep[LIRIS,][]{man98} 
at the WHT on 2011 March 23.  SDSS\,J1242 was observed in a continuous and dithered 
manner in each of the three filters with individual exposure times of 30\,s in $J$, and $H$, and 20\,s in $K_s$, for a total integration time of 540\,s 
in each filter.  Three standard star fields from the ARNICA catalogue \citep{hun98} were observed for flux calibration.  
The median sky subtracted frames were shifted and average combined into a single 
image per filter on which aperture photometry was performed with IRAF tasks.  
The aperture radii used for the flux standard were $r=3\farcs75$ with $5\farcs0$--$7\farcs5$ 
sky annuli, while smaller apertures of $1\farcs25$ were used for the science target, and corrected using several bright stars in 
each reduced frame.  The absolute flux calibration was good to 5~per cent or better, and the resulting $JHK_s$ photometry is listed in Table \ref{t:photometry}.

\subsection{{\em Spitzer} IRAC observations}

SDSS\,J1242 was observed on 2013 June 26 with the {\em Spitzer} warm mission \citep{wer04} as part of Cycle 9 program 90121.  Images 
of the science target were taken using the Infrared Array Camera (IRAC; \citealt{faz04}) at both 3.6 and 4.5\,$\mu$m, in a series 
of 30\,s exposures with 40 medium size dithers in the cycling pattern, for a total exposure time in each channel of 1200\,s.  Fully 
reduced, combined, and flux-calibrated images were analysed following \citet{Farihi10b} using $0.6$\,pixel$^{-1}$ mosaics created 
using MOPEX.  Aperture photometry was performed with tasks in IRAF using $r=4$ pixel radii and $24-40$ pixel sky annuli, 
applying aperture corrections recommended in the IRAC Instrument Handbook.  The IRAC photometry is listed in Table \ref{t:photometry} 
and shown in Figure \ref{f:model_sed}, where errors include the standard deviation in the sky annuli and an additional 5~per cent absolute calibration 
uncertainty.

\begin{table}
\begin{center}
\caption{Multi-wavelength photometry for SDSS\,J1242.}
\label{t:photometry}
\begin{tabular}{@{}lcD{.}{.}{8}D{,}{}{1}@{}}
\hline

Source		&$\lambda_{\rm eff}$	& \multicolumn{1}{c}{$m$}		& \multicolumn{1}{c}{$F_{\nu}$}\\
		&[$\mu$m]		& \multicolumn{1}{c}{[(AB) mag]}	& \multicolumn{1}{c}{[$\mu$Jy]}\\
\hline

{\em GALEX}	&0.15		& 21.0\,\pm\,0.3	      & 14\,\pm\,3\,\,\,\,\,\,\,\,\,,\\
{\em GALEX}	&0.23		& 18.91\,\pm\,0.06	      & 99\,\pm\,10\,\,\,\,\,\,,\\
SDSS $u$	&0.36		& 17.87\,\pm\,0.03	      & 270\,\pm\,14\,\,\,\,\,\,,\\
SDSS $g$	&0.47		& 17.74\,\pm\,0.02	      & 291\,\pm\,15\,\,\,\,\,\,,\\
SDSS $r$	&0.62		& 17.96\,\pm\,0.02	      & 238\,\pm\,12\,\,\,\,\,\,,\\
SDSS $i$	&0.75		& 18.19\,\pm\,0.02	      & 191\,\pm\,10\,\,\,\,\,\,,\\
SDSS $z$	&0.89		& 18.38\,\pm\,0.04	      & 157\,\pm\,8\,\,\,\,\,\,\,\,\,,\\
LIRIS $J$	&1.24		& 18.15\,\pm\,0.05	      & 87\,\pm\,4\,\,\,\,\,\,\,\,\,,\\
LIRIS $H$	&1.66		& 18.12\,\pm\,0.05	      & 58\,\pm\,3\,\,\,\,\,\,\,\,\,,\\
LIRIS $K_s$	&2.16		& 18.06\,\pm\,0.05	      & 40\,\pm\,2\,\,\,\,\,\,\,\,\,,\\
IRAC		&3.55		& 18.15\,\pm\,0.06	      & 15.5\,\pm\,0.7,\\
IRAC		&4.49		& 18.25\,\pm\,0.06	      &  9.0\,\pm\,0.5,\\

\hline
\end{tabular}
\end{center}
\end{table}
\section{Spectral and energy distribution analysis}
\label{chap3}
\subsection{Spectral modelling}
\label{chap3.1}
A first determination of the stellar parameters was obtained from the SDSS spectrum
and photometry.
The input physics of the stellar models that we used has been described extensively in \citet{Koester10}.
Since both hydrogen and helium lines are detected in the spectrum, we prepared a grid of models
with different H/He abundances. These models include both Stark broadening by electrons and He$^{+}$ ions,
and neutral broadening by He atoms. Due to the intrinsic difficulty in measuring the effect of surface gravity on spectral lines
for helium-rich atmospheres in this range of $T_{\rm{eff}}$, we initially fixed $\log{g} = 8$,
which corresponds to a typical 0.59\,$M_{\odot}$ white dwarf \citep[e.g.][and references therein]{Tremblay13}.
The best-fit temperature was 12\,500\,K. 
We also obtained initial abundance determinations for the elements that are visible in the SDSS spectrum:
O, Na, Mg, Si, Ca, and Fe.
\begin{figure*}
\centering
\includegraphics[width=0.9\linewidth]{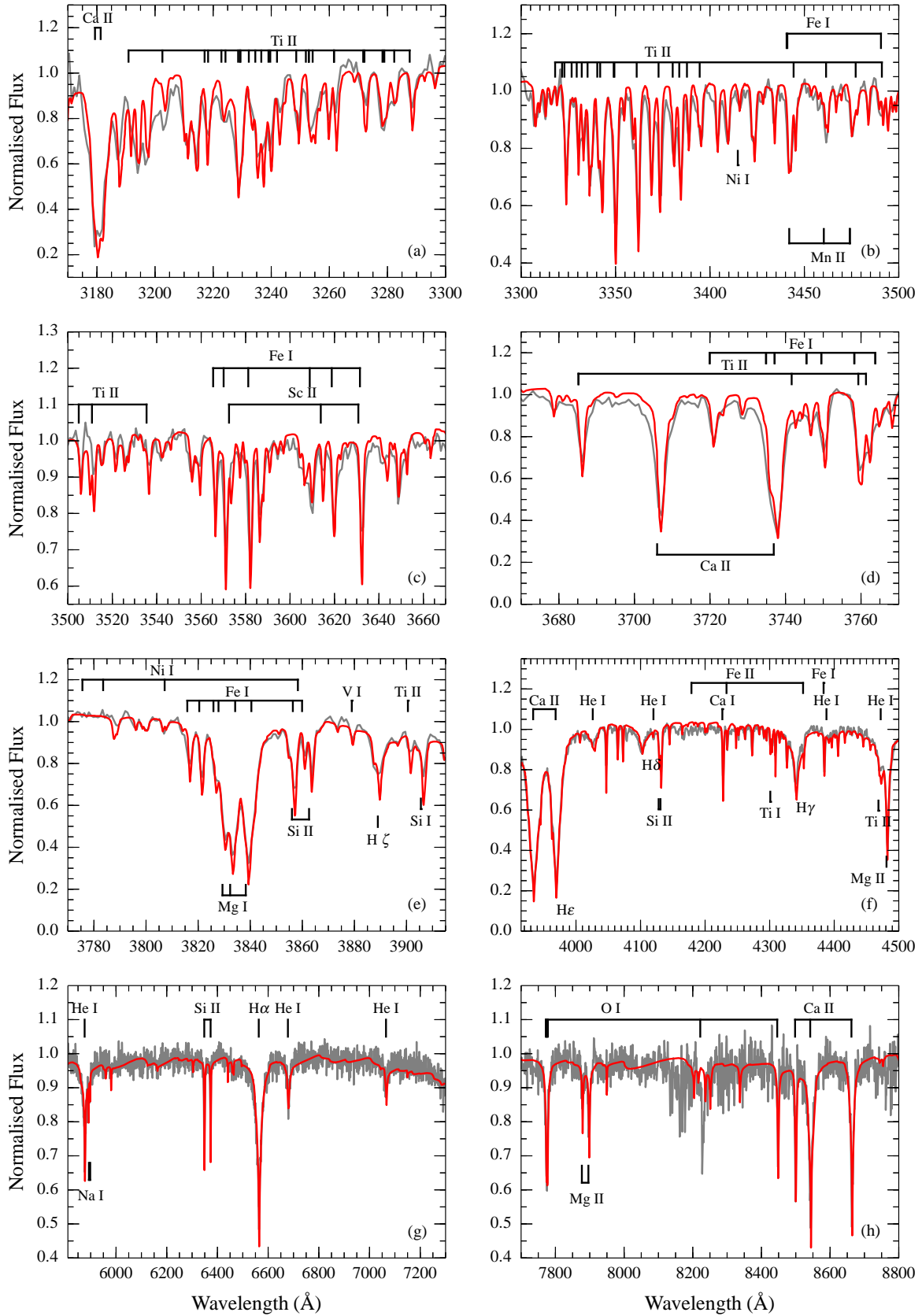}
\caption{WHT/ISIS spectrum (grey) and best-fit model atmosphere (red) with $T_{\rm{eff}}= 13\,000$\,K and $\log{g} = 8$. 
In panel (h), a problematic sky subtraction causes the noisier appearance of the spectrum at 8100--8300\,\AA. }
\label{f:isis_spectrum}
\end{figure*}

The best-fitting model was subsequently compared to the WHT/ISIS spectra
that are suitable for a more detailed chemical analysis.
We first noted that $T_{\rm{eff}}= 13\,000 \pm 300$\,K better reproduces the ionisation equilibria for the observed elements.
Next, we adjusted the initial abundances in steps of 0.1\,dex in order to improve individual abundances 
and error estimates for elements that have well-isolated lines \citep[verified by visual assessment as in][]{Gaensicke12}. 
This was possible for H, O, Na, Mg, Si, and Ca. The abundances of Ti, Cr, and Fe, 
which have many overlapping lines in the 3200--3700\,\AA\
range, were determined globally using regions where always one of the three elements is dominant. 

The WHT/ISIS spectrum along with the best-fit model are displayed in Fig.~\ref{f:isis_spectrum}. We note a complex of
Ti\,{\sc ii} lines in the range 3300--3800\,\AA, from which we determine precise abundances for this element. 
Fe\,{\sc i}, Fe\,{\sc ii}, Mn\,{\sc i}, and strong ultraviolet Ca\,{\sc ii} lines, in addition to Ca\,{\sc ii} H~\&\,K,
are also evident in the blue spectrum.
Other strong absorption lines that were already detected in the SDSS 
spectrum appear much more structured at the higher resolution and better signal-to-noise ratio of the WHT/ISIS data, in particular
Na\,{\sc i} D, Si\,{\sc ii}, O\,{\sc i}, Mg\,{\sc ii}, and Ca\,{\sc ii} triplet in the red spectra.
For a comprehensive list of the most important transitions, which are labelled Fig.~\ref{f:isis_spectrum}, see also \citet{Klein11}.

Only upper limits could be determined for C, N, Al, P, S, Sc, V, Mn, and Ni. Since these limits 
were in all cases larger than their solar abundances,
we included them in the model calculations fixed to solar ratios and verified
that the inclusion did not have significant influence on the fit.
The final abundances were then determined
statistically, using the standard width of the distribution from fits to different lines as error
estimate, and are summarised in Table~\ref{t:abd_ISIS}.
\begin{table}
\centering
\caption{Photospheric element abundances derived from the WHT/ISIS spectrum,
adopting $T_{\rm{eff}}=13\,000$\,K, and $\log{g} = 8$.} 
\label{t:abd_ISIS}
\begin{tabular}{@{}rlD{,}{\,\pm\,}{4}@{}}
\hline
Z & Element & \multicolumn{1}{c}{[Z/He]}\\
\hline
 1 & 	 H  &  -3.68  , 0.10  \\   
 6 & 	 C  & <-4.70     \\	   
 7 & 	 N  & <-5.00     \\	   
 8 & 	 O  &  -4.30 , 0.10   \\   
11 & 	Na  &  -7.20 , 0.20   \\   
12 & 	Mg  &  -5.26  , 0.15  \\   
13 & 	Al  & <-6.50     \\	   
14 & 	Si  &  -5.30  , 0.06  \\   
15 & 	 P  & <-6.60     \\	   
16 &     S  & <-8.00     \\	   
20 & 	Ca  &  -6.53  , 0.10  \\   
21 & 	Sc  & <-9.50     \\	   
22 & 	Ti  &  -8.20 , 0.20   \\   
23 & 	 V  & <-9.00     \\	   
24 & 	Cr  &  -7.50 , 0.20   \\   
25 & 	Mn  & <-8.00     \\	   
26 & 	Fe  &  -5.90 , 0.15  \\     
28 & 	Ni  & <-7.30     \\	   
\hline
\end{tabular}
\end{table}

From the WHT/ISIS spectrum we also have estimated how a different surface gravity influences the measurement 
of the metal abundances. In fact, the ionisation balance is affected by the surface gravity and is coupled to the electron density, 
while the line strengths depends also on the neutral density via the line broadening. 
Helium-dominated white dwarfs are suspected to have masses in the range 0.67--0.74\,$M_\odot$ \citep[][]{Bergeron11,Falcon12}, which correspond to
slightly larger surface gravities than the canonical $\log{g} = 8$ that we adopted.  
However, we cannot exclude SDSS\,J1242 to be less massive than 0.60\,$M_\odot$, therefore 
we altered $\log{g}$ by $\pm 0.25$\,dex, which spans the mass range of most known white dwarfs \citep[see][]{Tremblay13},
i.e. $0.45$--$0.80 M_{\odot}$. The unknown surface gravity induces systematic uncertainties of 
0.02--0.07\,dex for H, Na, Mg, and Fe, and up to 0.2\,dex for O. 
The effect that different surface gravities have on the relative abundance ratios is discussed in Section~\ref{chap4.1}, 
where we consider the diffusion of elements in the convection zone of SDSS\,J1242.

We have estimated the interstellar reddening $E(B-V)$, measuring the colour excess in the optical. This is
computed with respect to the synthetic colours determined from the best-fit model, which are: 
$(u-g) = -0.041$, $(g-r) = -0.238$, $(r-i) = -0.238$, and $(i-z) = -0.286$. We obtain $E(B-V) = 0.05 \pm 0.02$, which agrees within 2$\sigma$ with the 
total Galactic reddening measured along the line-of-sight to SDSS\,J1242 \citep[$E(B-V) \approx 0.015$,][]{SFD98}.
We determine a spectroscopic parallax of $164 \pm 28$~pc, using the absolute $g$ magnitude scale for DB white dwarfs
by \citet{Bergeron11} and $A(g) =  3.77 \times E(B-V) = 0.19 $ from the standard \citet[][]{Cardelli89} reddening law.

\subsection{Search for dust emission in the infrared}
\label{chap3.2}
\begin{figure}
\centering
\includegraphics[width=\linewidth]{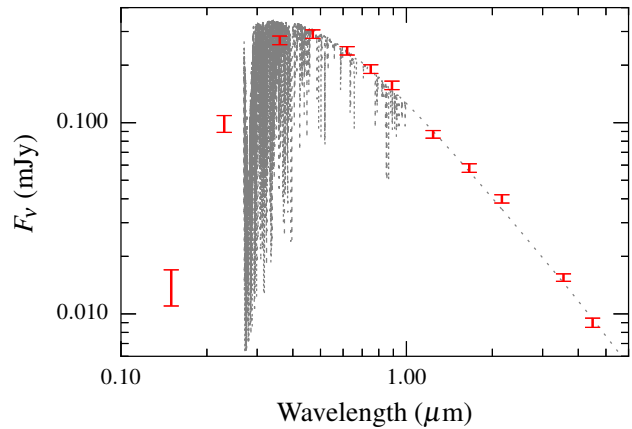}
\caption{The model atmosphere with $T_{\rm{eff}} = 13\,000$\,K, $\log{g} = 8$ is shown. The error bars represent
the broad-band photometry given in Table~\ref{t:photometry}.}
\label{f:model_sed}
\end{figure}
In Fig.~\ref{f:model_sed} we display all available photometry for SDSS\,J1242, including the ground- and space-based data from the ultraviolet through the infrared.  
The data are overplotted with the best fitting atmospheric model as described in the previous section.  

It is noteworthy that the $K_s$-band data point lies about 2$\sigma$ above the predicted photosphere, and the $JHK_s$ colours appear redder than the model slope.
The flux calibration for the $JHK_s$ photometry was scrutinised for possible sources of error, 
and while none were found, there was only a single, faint 2MASS point source in the LIRIS field 
for an independent zero point determination.  Given the IRAC fluxes, which are both self consistent 
and relatively robust in accuracy, the conclusion must be the $JHK_s$ fluxes contain some, modest additional error.

As can be seen from the figure, relative to the model there is no excess infrared emission that might originate in circumstellar dust.
While an in-depth discussion of the detectability of discs at white dwarfs is beyond the scope of this paper, 
because the accretion history of SDSS\,J1242 influences the forthcoming interpretation of the observed heavy elements, 
a brief discussion is useful here.  First and most relevant is that narrow rings of solid material holding up to $10^{22}$\,g 
can remain undetected at white dwarfs for inclinations above 50 degrees \citep[][]{Farihi10b}.  
Such rings can account for the highest instantaneous accretion rates, and for the bulk of parent body masses inferred for polluted white dwarfs.  
Second, a population of narrow rings is indicated by recent {\em Spitzer} studies that reveal a growing number of subtle infrared excesses \citep[][]{Bergfors14}, 
and where the frequency of infrared excesses rises sharply at fainter disc luminosities \citep[][]{Rocchetto14}.

The implications of the non detection of infrared emission from circumstellar dust become relevant in the following section, where we discuss the likely
accretion history for SDSS\,J1242 and detail the atmospheric composition.

\section{Discussion}
\label{chap4}
\subsection{Diffusion analysis}
\label{chap4.1}
White dwarfs with helium-rich (DB) atmospheres develop a convection zone for
$T_{\rm{eff}} \lesssim 25\,000$\,K \citep[see fig.~3 in][]{Bergeron11}. The depth of the convection zone, 
typically expressed as a fraction of the white dwarf mass, 
$q_{\rm{cvz}} = \log{(M_{\rm{cvz}}/M)}$, depends on the atmospheric parameters $T_{\rm{eff}}$ and $\log{g}$. 
For SDSS\,J1242, we infer $q_{\rm{cvz}} = -5.4$ using the most up-to-date version of the stellar structure computations described in \citet{Koester09}, which
adopt the mixing-length approximation \citep{Bohm58} with a parameter ML2/$\alpha = 1.25$.
White dwarf mass, radius, and age are estimated using the DB cooling models of the 
Montreal group\footnote{Available at: \url{http://www.astro.umontreal.ca/~bergeron/CoolingModels/}}, for $\log{g} = 8.00 \pm 0.25$, 
and are given in Table~\ref{t:parameters_cvz} along with the values of $q_{\rm{cvz}}$.
\begin{table}
\centering
\caption{Physical parameters of SDSS\,J1242.}
\label{t:parameters_cvz} 
\begin{tabular}{@{}lrrr@{}}
\hline
\multirow{2}{*}{Parameters}     & \multicolumn{3}{c}{$T_{\rm{eff}} = 13\,000 \pm 300$\,K}\\
                & $\log{g} = 7.75$ &  $\log{g} = 8.00$ & $\log{g} = 8.25$ \\
\hline
Mass      $(M_{\odot})$  & 0.48      & 0.59  & 0.75       \\
Radius    $(R_{\odot})$  & 0.015     & 0.013 & 0.011       \\
Cooling age (Gyr)         & 0.25      &  0.32 & 0.50                 \\
Distance    (pc)         & 193       &  164  & 137\\
$q_{\rm{cvz}}$                      & $-4.9$    & $-5.4$  & $-5.9$ \\
$\log{M_{\rm{H}}}$ [g]    & 23.80  & 23.38 & 22.99\\
\hline
\end{tabular}
\end{table}
\begin{table}
\caption{Diffusion data for SDSS\,J1242 for the $T_{\rm{eff}} = 13\,000$\,K and $\log{g} = 8$ model.
From left to right: the mass of elements that is mixed in the convection zone, diffusion velocities,
diffusion time scales, and mass fluxes \citep[see][for more details]{Koester09}.} 
\label{t:abundances_cvz}
\centering
\begin{tabular}{@{}lD{.}{.}{5}rrD{.}{.}{2}@{}}
\hline
Element   &\multicolumn{1}{r}{$M_{\rm{Z}}$}&      $v$	      &   $\tau$	 & \multicolumn{1}{r}{$\dot{M_{\rm{Z}}}$}\\
          &\multicolumn{1}{r}{(10$^{23}$\,g)}&$10^{-8}$ (cm/s)&$$ (Myr)  &\multicolumn{1}{r}{$10^{7}$(g/s)} \\
\hline

    C   & 	<2.8	  &  3.5&    2.3&<  371.27 \\
    N   & 	<1.7	  &  3.7&    2.2&<  228.91 \\
    O   & 	 9.6	  &  3.6&    2.2&  1282.10 \\
   Na   & 	 0.02	  &  3.7&    2.2&     2.36 \\
   Mg   & 	 1.6	  &  3.3&    2.4&   197.35 \\
   Al   & 	<0.1	  &  3.6&    2.2&<   13.49 \\
   Si   & 	 1.7	  &  4.6&    1.7&   285.23 \\
    P   & 	<0.09	  &  6.9&    1.2&<   23.72 \\
    S   & 	<0.004    &  7.9&    1.0&<    1.13 \\
   Ca   & 	 0.1	  &  8.0&    1.0&    42.26 \\
   Sc   & 	<0.0002   &  8.6&    0.9&<    0.05 \\
   Ti   & 	 0.004    &  8.5&    0.9&     1.13 \\
    V   & 	<0.0006   &  8.4&    0.9&<    0.19 \\
   Cr   & 	 0.02	  &  7.9&    1.0&     5.72 \\
   Mn   & 	<0.006    &  7.8&    1.0&<    1.90 \\
   Fe   & 	 0.8	  &  7.4&    1.1&   228.85 \\
   Ni   & 	<0.03	  &  6.6&    1.2&<    8.80 \\
\hline						          
Sum*    &        12.4 &              &           & 2045.00 \\
\hline
\multicolumn{5}{l}{*: Upper limits are not included}\\
\end{tabular}
\end{table}

In the atmosphere of a convective white dwarf, elements heavier than hydrogen and helium 
will eventually diffuse out with time-scales that depend on the atomic weight of the element
 and the local conditions at the bottom of the convection zone.
Given the considerable length of these diffusion time-scales, $\approx 1$\,Myr in the case of SDSS\,J1242, variations in the
accretion rate on shorter time-scales can result in differences between the photospheric abundances,
and those of the accreted debris. \citet{Koester09} illustrates the accretion on to a white dwarf in a simple fashion, consisting of 
three phases: i) build-up or early-phase, during which the abundances in the photosphere increase and the relative abundance
ratios equal those of the debris; ii) steady-state, when photospheric and debris abundances scale with the relative diffusion velocities,
as the equilibrium between accretion and diffusion is reached;
iii) declining stage, when the accretion has stopped and the photospheric abundances decrease exponentially with different diffusion time-scales for each element.

Here, we consider the accretion-diffusion equilibrium, which is the only
well-modelled phase, providing a lower limit on the mass of debris that has been
accumulated in the convection zone.
From diffusion calculations that follow the prescriptions of \citet{Koester09}, we estimate for each detected element the diffusion time-scale, 
diffusion velocity, and the mass flux throughout the atmosphere -- which is equal to the time-averaged accretion rate on to the star. Since the diffusion data depend on 
the white dwarf mass, we investigate the effect that the uncertainty of $\Delta \log{g} = \pm 0.25$~dex has on the calculations. 
We note that a decrease (increase) of 0.25\,dex results in: a) 40 per cent decrement (increment) in the average diffusion velocities; b)
the mean diffusion time-scales become about 3 times longer (shorter); c) the mean mass fluxes that are required to
produce the observed traces of elements in the stellar atmosphere are 20 per cent smaller (larger). 
We conclude that the diffusion parameters scale uniformly with a change of surface gravity,
differing very little for individual elements, of the order of few per cent, from the average scaling.
Finally, the mass flux ratios also remain unchanged within a few per cent.
Changes induced by the temperature uncertainty are irrelevant in comparison. 
Therefore, in Table~\ref{t:abundances_cvz} we give the diffusion data
relative to $T_{\rm{eff}} = 13\,000$\,K and $\log{g} = 8$ only. 
\subsection{Accretion rate and parent body mass}
\label{chap4.2}

The diffusion analysis makes SDSS\,J1242 a remarkable object.
First, the mass of heavy elements within its convection zone is
 $1.2 \times 10^{24}$\,g, larger than the mass of Ceres, the largest asteroid in the Solar System 
which is best described as a fully differentiated, planetary embryo.
This is actually a lower limit to the total accreted mass, and makes this star one of the most polluted white dwarfs known. 
Second, the total accretion rate we derive (averaged over the diffusion time-scales of $\approx 1$\,Myr), of $2 \times 10^{10}$\,g\,s$^{-1}$, 
is also among the highest of all known metal-polluted white dwarfs \citep[][]{Girven12, Dufour12}.

In the case helium-dominated atmospheres with Myr-long diffusion time-scale, 
ongoing accretion is uncertain in the absence of an infrared excess associated with closely orbiting dust.  
Furthermore, while the time-averaged accretion rate of SDSS\,J1242 is set, its accretion history is unknown, and specifically 
whether the rate has changed over the last one to few metal sinking time-scales.  There is some observational evidence that accretion 
rates change over time, where relatively short bursts of high rates
may influence the time-averaged rates for stars like SDSS\,J1242 \citep[][]{Farihi12}.

In the absence of dust detection, it is difficult to construct a scenario in which the accretion on to SDSS\,J1242 is in the early phase 
(i.e. where a single diffusion time-scale has not yet passed since the onset).  This is because the implications would be either 1) 
the inferred accretion rate is also the instantaneous rate, and a disc with mass $\gtrsim 10^{24}$\,g would require an unlikely, near edge-on configuration
to escape detection, or 2) an entire planetary body exceeding the mass of Ceres was accreted in a short time ($< 1$\,Myr) 
and the system is observed at a special epoch.  
Given the potential for discs to linger for several Myr \citep[][]{Klein10,Bochkarev11,Girven12}, 
it is more likely SDSS\,J1242 is accreting now at a relatively modest rate compared to its time-averaged rate, and could have formed a disc 
that is too subtle to be detected in our {\em Spitzer} data, due to being gas-dominated, narrow, inclined, or a combination of these factors. 

Given the 1--2 Myr time-scales for heavy elements to diffuse from the convection zone, the star could be in a declining phase (as opposed to a steady state), 
where accretion halted at least one diffusion time-scale ago.  In that situation, the metal mass present in the outer layers of the star would be decreasing exponentially. 
If indeed several Myr have passed since the end of the accretion episode, the accreted parent body would have to be more massive than Pluto to account for
the currently observed metal abundances.
Because the accretion of entire planets is expected to happen only rarely \citep[][]{Veras13}, such events are
unlikely to be detected and a significant passage into a declining phase of accretion can be discounted.

\subsection{Composition of the debris}
\label{chap4.3}

Analysing in detail the atmospheric composition of SDSS\,J1242, we note that oxygen, magnesium, silicon, and iron
\citep[which alone make up 94 per cent of bulk Earth,][]{McDonough01},
account for 99 per cent by mass of metals we detect. 
In particular, oxygen is the most abundant element in the photosphere of SDSS\,J1242, with a mass fraction of $64 \pm 13$ 
per cent. Oxygen is followed by magnesium, silicon, and iron
with $10\pm 1$, $14 \pm 1$, and $11 \pm 2$ per cent respectively. SDSS\,J1242 has an iron and magnesium content
that is compatible with basaltic rocks typical of a wide array of
achondritic meteorites, such as howardites, eucrites, and diogenites \citep[see fig.\,6 in][]{Nittler04}.
These compositions result from partial melt or fractional crystallisation of silicate rocks, inducing magnesium enhancement,
and are observed on the surface of differentiated asteroids in the Solar System \citep[e.g. Vesta;][]{DeSanctis12}. 

To compare our results to those concerning the bulk composition of Solar System meteorites,
we compute the relative weight ratios of magnesium, silicon, and iron. We note that, in the steady state, $\rm{Mg/Si} = 0.70$ and 
$\rm{Fe/Si} = 0. 80$ agree with those of angrites, which are meteorites resulting from the fractional crystallisation of silicates 
due to impacts \citep[see fig.\,7 in][]{Nittler04}. We also find that $\rm{Cr/Fe} = 0.025$ is within Solar System 
\citep[typically between 0.01--0.03;][]{Nittler04} and extrasolar asteroids values \citep[e.g.][]{Jura14}, 
in agreement with the hypothesis that siderophile refractory elements would condensate along with iron in rocks.
The relatively low iron abundance is consistent with accretion of rocks from a thick crust and mantle of
a large asteroid. Given their shorter diffusion time-scales, it may be possible that atmospheric abundances of heavier core materials 
(iron and nickel) were larger in the past, and these elements have now mostly diffused out of the convection zone.
However, as discussed in Section~\ref{chap4.2} and in the following section, we argue against a long time since the end of accretion (more
than a few diffusion time-scales) that would be needed to drastically change the relative abundances of metals.

In conclusion, our results suggest an intricate history for the parent body that polluted SDSS\,J1242, which probably underwent 
geological evolution, possibly caused by impacts with other asteroids.
These may have caused melting and mixing of the crust, leaving a chemical signature in its composition that resemble findings for
Solar System asteroids.

\subsection{Oxygen excess}
\label{chap4.4}
\begin{table}
\caption{Oxygen mass fraction balance in SDSS\,J1242}
\label{t:oexcess}
\centering
\begin{tabular}{@{}lD{.}{.}{12}@{}}
\hline
Oxide& \multicolumn{1}{r}{Mass Fraction}\\
\hline
Na$_2$O      &    0.0006\,\pm\,0.0002\\
MgO          &    0.10\,\pm\,0.02\\
Al$_2$O$_3$  &< 0.009      \\
SiO$_2$      &    0.25\,\pm\,0.05\\
CaO          &    0.013\,\pm\,0.004\\
TiO$_2$      &    0.0006\,\pm\,0.0002\\
Cr$_2$O$_3$  &    0.0021\,\pm\,0.0005\\
FeO          &    0.05\,\pm\,0.01\\
\hline
O excess     &  0.57\,\pm\,0.07\\
H$_2$O       &  0.38\,\pm\,0.10\\
\end{tabular}
\end{table}

Since we detect all four major rock forming elements, plus minor constituents, we can assess the 
fraction of oxygen that was originally bound on anhydrous minerals within the parent body. 
We can safely exclude that the oxygen observed in the convection zone is dredged-up core-material, 
since all the white dwarfs that have dredged-up oxygen also show optical C\,{\sc i} lines \citep[see e.g.][]{Gaensicke10} that we do not detect.
Therefore, we ignore carbon in the following discussion, but comment on its possible impact at the end of this section.

We follow \citet{Klein10, Klein11} and \citet{Farihi11, Farihi13}, 
adopting the fundamental metal-to-oxygen combinations found in
rocks, i.e. Na$_2$O, MgO, Al$_2$O$_3$, SiO$_2$, CaO, TiO$_2$, Cr$_2$O$_3$, and FeO. 
Iron can also be metallic or rarely form Fe$_2$O$_3$, but we note that including 
only FeO in the computation is a conservative assumption, and an upper limit to the
number of oxygen atoms that are bound to iron. Our results are reported in Table~\ref{t:oexcess}. 
The errors are determined via a Monte Carlo method, 
with Gaussian probability distributions whose widths are set by the abundance uncertainties. 
Under the assumption of steady state, we find 57 per cent of oxygen mass to be in excess with respect to that 
needed to form simple anhydrous metal oxides. Because water is commonly found
in asteroids, the natural explanation for this oxygen excess
is that the accreted debris was rich in water ice or hydrated minerals. 
Taking into account the mass of metals in the convection zone of SDSS\,J1242, we estimate that the 
accreted parent body was made of 38 per cent H$_2$O by mass.
We also note that $\rm{O/Si} = 4.5$ is equivalent to the most extreme ratio measured for hydrated silicates that are found in CI and CM chondrites,
which are rich in water and suspected to form in proximity of the Solar System snow line. This is a hint
that the surplus oxygen may have not been carried only by rocks, but might have been bound in water ice.

\begin{figure}
\centering
\includegraphics[width=\linewidth]{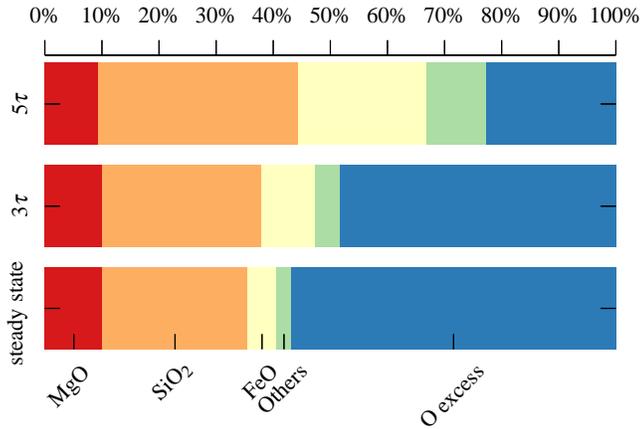}
\caption{Oxygen balance within metal oxides at the steady state, and computed under the assumption of declining phase 
after 3 and 5 time-scales ($\tau \approx 1$\,Myr) from the halt of the accretion. 
The atmospheric abundances are assumed to decrease exponentially with time. The declining phase scenario implies smaller initial oxygen excess. 
We now measure $M_{\rm{Z}} = 1.2 \times 10^{24}$\,g, i.e. 1.4 Ceres masses, while 
 we would expect SDSS\,J1242 to have accreted at least $6\times10^{24}$\,g (7 Ceres masses), and $2 \times 10^{25}$\,g (1.5 Pluto masses),
 in the case we observe the white dwarf at $3\tau$ and $5\tau$ since the end of the accretion.}
\label{f:oexcess}
\end{figure}

If we considered instead that SDSS\,J1242 is observed in the decline phase, the oxygen abundance relative to other
elements would change with time. Most metals that combine with oxygen diffuse faster out of the convection zone, 
and would have been more abundant at the end of the accretion phase. We compare in Fig. \ref{f:oexcess} the
steady state oxygen balance with what could have been the parent body composition, placing the end of accretion at a few 
to several diffusion times in the past. We note a reduction of the oxygen excess with respect to fraction bound in metal oxides.
In other words, if sufficient time has passed since the end of the accretion phase, even a dry asteroid
would produce an apparent oxygen excess. However, as discussed in Section~\ref{chap4.2}, we argue against a scenario
in which SDSS\,J1242 is observed late in to the declining phase. 
Furthermore, the remarkable trace hydrogen content (Section\,\ref{chap4.5}) of this helium-dominated 
white dwarf would require an additional mechanism beyond the extant pollution event.

Finally, we consider how the inclusion of solar-abundance carbon in the oxygen balance would affect our result.
If carbon were found to be polluting SDSS\,J1242 at the upper limit, 
then all of the excess oxygen can be accounted for as CO$_2$.
However, this assumption is invalid because our optical limits are not very stringent. In the eight cases for which metal-polluted 
white dwarfs have been observed in the ultraviolet, their carbon abundances have been found 
to be at least 1000 times less abundant than the solar value \citep[][]{Gaensicke12,Jura12a,Xu14}.
Furthermore, only two metal-polluted white dwarf display a large abundance of
carbon, the origin of which is still debated (PG\,1225$-$079 in \citealt{Xu13}, 
and Ton\,345 in \citealt{Jura15}; Wilson et al. submitted).

\subsection{Hydrogen mass}
\label{chap4.5}
\begin{figure*}
\centering
\includegraphics[width=\linewidth]{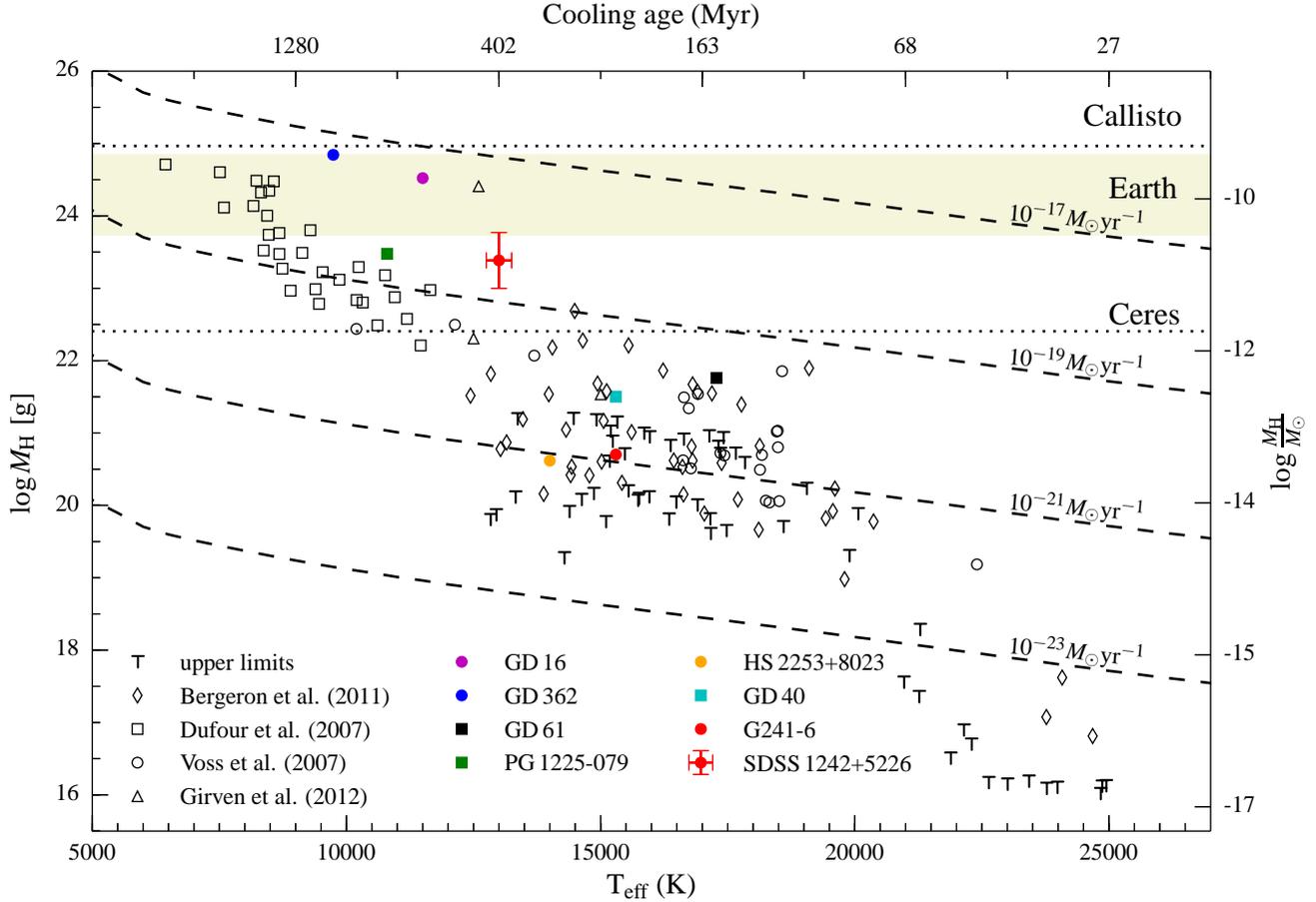}
\caption{Total mass of hydrogen in the convection zones of helium-dominated white dwarfs as function of $T_{\rm{eff}}$ (left axis). 
On the right axis, the hydrogen mass is given in units of solar masses and we give the cooling age for a typical 
$\log{g} = 8 $ DB white dwarf \citep[][]{Bergeron11} on the top axis. Hydrogen abundances of field DBA and DZA white dwarfs (open symbols, see legend for references)
were converted into hydrogen mass adopting the evolutionary models of \citet{Bergeron11}, and $q_{\rm{cvz}}$
from the updated version of \citet{Koester09} computations. The hydrogen abundances of metal-polluted white dwarfs
are taken from \citet[][and references therein]{Jura14}. For GD\,362 \citep[][]{Gianninas04} and GD\,61 \citep[][]{Farihi13} 
we used the hydrogen mass supplied by the authors. The mass andof hydrogen in water for Ceres, Callisto 
\citep[ice mass fractions are 25 and 50 per cent respectively,][]{Michalak00,McCord05,Canup02}
and bulk Earth \citep[][and references therein]{vanThienen07} are plotted as dashed lines or the shaded area.
The dashed curves represent the cumulative mass of hydrogen for constant accretion rates during the white dwarf cooling ages.}
\label{f:hydrogen_mass}
\end{figure*}

Assuming that hydrogen is uniformly mixed within the convection zone of SDSS\,J1242, 
the large abundance we measure, [H/He]$ = -3.68$,  corresponds to
$M_{\rm{H}} = 2.5 \times 10^{23}$\,g. 
Under the assumption of steady state, we have inferred that H$_2$O 
accounts for about 38\,per cent of the parent-body mass, meaning that
an asteroid of about $5.9\times 10^{24}$\,g is needed to yield all observed $M_{\rm{H}}$.
The minimum parent body mass of $1.24\times 10^{24}$ implies that at least $5.2 \times 10^{22}$\,g
of H was added to the envelope of SDSS\,J1242 in the observed
accretion episode, or 20\,per cent of the total $M_{\rm{H}}$. It is possible that the
entire $M_{\rm{H}}$ was accreted in this current episode, requiring steady
state at the measured average rate for a few Myr.  However, it
cannot be excluded that multiple accretion episodes of water-rich
bodies contributed to the total H-content of the envelope over the
300\,Myr cooling age of the star.

We compare the $M_{\rm{H}}$ of SDSS\,J1242 with that of other helium-rich white dwarfs in Fig.~\ref{f:hydrogen_mass}, which has
also been represented in different fashion and discussed in various contexts \citep[e.g.][]{MacDonald91,Voss07,Dufour07,Bergeron11}.
We note that SDSS\,J1242 holds at least one order of magnitude more hydrogen than stars with similar temperatures (i.e. cooling ages). 
Several hypotheses have been put forward to explain the apparent broad correlation between the hydrogen content of helium-rich white
dwarfs and their effective temperature ($\equiv$ cooling age): i) spectral type evolution, 
which requires the presence of small amounts of primordial hydrogen 
($\lesssim 10^{-14} M_{\odot}$) that gets mixed in the helium-rich envelope once convection sets in \citep[][]{MacDonald91,Bergeron11}; ii) accretion 
from the interstellar medium \citep[ISM;][]{Dupuis93a,Dufour07}; iii) accretion of  
water-rich planetary debris \citep[][]{Jura09a,Farihi10a}.
The primordial-hydrogen hypothesis can provide the observed $M_{\rm{H}}$ of $T_{\rm{eff}} \gtrsim 20\,000$\,K DBA white dwarfs,
but fails to explain the hydrogen content of $T_{\rm{eff}} \lesssim 15\,000$\,K atmospheres. The lack of hot DBA white dwarfs
with large $M_{\rm{H}}$ requires an external origin for the hydrogen.
\citet{Farihi10a} strongly argued against the ISM accretion scenario, using several distinct lines of evidence, 
including Galactic positions and kinematics, ISM and stellar chemistry, and likely physical accretion mechanisms.
This also suggested that some DAZ white dwarfs are externally polluted by both metals and hydrogen. 

Here we discuss additional evidence to the planetary interpretation using Fig.~\ref{f:hydrogen_mass}.
We note that the $M_{\rm{H}}$ detected in helium-rich
atmospheres spans 8--9 orders of magnitude and, if interpreted as continuous accretion from the ISM, 
it requires the accretion rates to be unlikely higher in old white dwarfs (dashed curves). 
While the observations of metal-polluted white dwarfs mainly suggests accretion of dry planetary debris 
\citep[$\leq 10$ per cent H$_2$O by mass;][]{Klein11,Jura12b},
pollution from water-rich debris seems plausible in a handful of stars with extremely large $M_{\rm{H}}$
\citep[GD\,16, and GD\,362, see][]{Jura09a,Jura12b}, and has been confirmed
in GD\,61 \citep[][]{Farihi13} and likely in SDSS\,J1242.
We note that GD\,16, GD\,362 and SDSS\,J1242 are extremely metal polluted, and the first two along with GD\,61 display near-infrared excess, 
revealing the presence of a debris disc that unambiguously confirms ongoing accretion. 
We also note that a further star possessing a large $M_{\rm{H}} \approx 5 \times 10^{24}$\,g
is HE\,0446$-$2531 with $T_{\rm eff} = 12\,600$\,K. However, its spectral analysis 
is obtained from low-resolution spectra \citep{Friedrich00}, and it is found not to have 
circumstellar disc by \citet{Girven11}. 
Finally, we note that about 45 per cent of the 108 helium-dominated white dwarfs analysed by \citet{Bergeron11} show trace hydrogen, 
similar to the fraction of metal-polluted white dwarfs \citep[20--30 per cent;][]{Zuckerman03, Zuckerman10,Koester14}.
We therefore suggest that a large fraction (maybe most) of the helium-rich white dwarfs with trace hydrogen, 
but no signs of metal pollution, may have accreted water-rich debris in the past.
The metals eventually diffuse out the convection zone once accretion ceases,
while hydrogen remains in the outer, photospheric layers. Several diffusion time-scales after the end of an accretion episode, 
stars like GD\,61, which has only a modest $M_{\rm{H}}$ in Fig.~\ref{f:hydrogen_mass}, will resemble a typical DBA 
white dwarf, while SDSS\,J1242, GD\,16, and GD\,362 will resemble other 
helium-dominated white dwarfs with similar temperatures, but the larger $M_{\rm{H}}$ would be the only observable relic 
of their past accretion.

It is likely that extra-solar asteroids and planets, like those in our own Solar System, display a variety of compositions and water mass fractions.
Water accounts for a large mass fraction of asteroids and solar-system moons \citep[][]{Potsberg11,Roth14}, 
but it is a minor constituent of inner solar-system bodies.  
This characteristic diversity is thought to be common place in planet formation \citep[][]{Bond10} and, in the Solar System, 
there is evidence for an overall gradient of water content within rocky planetary bodies, increasing from the inner to the outer Solar System.  
The amount of water in asteroids and rocky planets is mostly determined by the initial (local) conditions constraining the snow line
in the protoplanetary disc, where water condenses into ice, but it is also affected by subsequent interactions between planetary embryos \citep[][]{Walsh11, Morbidelli12}.
It appears that ice may survive in asteroids orbiting white dwarfs, if they are distant enough from the star to avoid engulfment 
during the giant branches \citep{Mustill12} and large enough to shield the internal water from the increased stellar luminosity \citep{Jura10}.
Alternatively, water can survive in hydrated minerals that are significantly more resistant to heating. 
Also comets may be suggested to contribute to the fraction of $M_{\rm{H}}$ in helium-rich white dwarfs, but their 
delivery to the surfaces of white dwarfs should be less frequent, and have not yet been observed \citep[][]{Veras14b}.
\subsection{Summary}

The signature of water accretion on to white dwarf can exhibit as an oxygen excess \citep[][]{Jura10}, 
and has been unambiguously confirmed in GD\,61 \citep[][]{Farihi13}, and likely for SDSS\,J1242.
The detailed composition analysis of SDSS\,J1242 has reinforced the case for water accretion in other similar white dwarfs,
and possibly how that water is delivered (potentially as ice in this case).
The evidence points to an asteroid once orbiting an early type main sequence star that has possibly experienced a series of impacts. 
We have compared its composition to that of Solar System meteorites, suggesting that the crust of the asteroid underwent geological evolution.
From the magnesium and iron abundances, we suggested that melting and crystallisation of
the rocks at the surface may have taken place, likely following collisions with other asteroids. 
Large asteroids like the one that polluted SDSS\,J1242, likely comparable in size or larger than Ceres,
are suggested to have thick crusts made of silicates. While silicates could be important carriers of water, 
especially within the interior asteroids formed near or beyond the snow line,
thick layers of ice are suspected to hide underneath the surface of outer main belt asteroids such as Ceres.
It is therefore plausible that most of the oxygen in SDSS\,J1242 could have been water ice, 
and sufficiently deep within the parent body to prevent evaporation during the giant branches.

We stress that the discovery of stars showing oxygen excess establishes a link between the trace hydrogen observed in helium-dominated white dwarfs
to the accretion of water.
Studies targeting more stars with detailed abundance analysis will be able to reveal the frequency of water-rich asteroids.

\section{Conclusions}
\label{chap5}
We have presented the discovery and analysis of a new, strongly metal-polluted white dwarf. 
We detect trace-hydrogen and eight metal species in its helium-dominated atmosphere, which we interpret
as being accreted from planetary debris. 
The four most abundant elements are the major rock forming metals 
O, Mg, Si, and Fe, making up 99 per cent of the metals in the convection zone. 
With over $10^{24}$\,g of metals in its outer layers and an inferred accretion rate over
$10^{10}$\,g\,s$^{-1}$, SDSS\,J1242 is one of the most polluted white dwarfs known.
Its convection zone contains $2.5 \times 10^{23}$\,g of hydrogen,
which is about an order of magnitude larger than detected in helium-dominated white dwarfs of
similar temperature. From the analysis of the polluting debris, we conclude that SDSS\,J1242 likely accreted the remains of
a water-rich planetesimal, with an H$_2$O mass fraction of roughly 38 per cent.
 
With SDSS\,J1242, we reinforce the suggestion advanced by previous work on other strongly metal-polluted white
dwarfs. These hydrogen- and oxygen-enriched stars are the metaphorical ``tip of an iceberg'' of water-accreting white dwarfs.
A significant fraction of helium-dominated white dwarfs
could accrete water-rich debris, which, accumulated over their long cooling ages,
may account for the observed correlation between $M_{\rm{H}}$ and cooling age.

\section{Acknowledgements}
The WHT is operated on the island of La Palma by the Isaac Newton Group in the 
Spanish Observatorio del Roque de los Muchachos of the Instituto de Astrofísica de Canarias. 
The ISIS spectroscopy was obtained as part of SW2012a11 and W12AN005.
LIRIS photometry was obtained as part of WHT/2011A/12.
This work is based in part on observations made with the {\em Spitzer Space Telescope} 
(for program 90121), which is operated by the Jet Propulsion Laboratory, California Institute of Technology under a contract with NASA

We would like to thank Dimitri Veras for the useful discussion and for sharing the latest
results of his work.

The research leading to these results has received funding from the
European Research Council under the European Union's Seventh Framework
Programme (FP/2007-2013) / ERC Grant Agreement n. 320964 (WDTracer). 

Funding for SDSS-III has been provided by the Alfred P. Sloan
Foundation, the Participating Institutions, the National Science
Foundation, and the U.S. Department of Energy Office of Science. The
SDSS-III web site is http://www.sdss3.org/.

\bibliographystyle{mn2e}

\bibpunct{(}{)}{,}{a}{}{;}

\label{lastpage}

\end{document}